\documentclass{aastex6}

\usepackage{amsmath}
\makeatletter
\def\env@matrix{\hskip -\arraycolsep % taken from amsmath.sty lines 895ff
  \let\@ifnextchar\new@ifnextchar
  \array{*{\c@MaxMatrixCols}c}}
\makeatother

\usepackage{amssymb}
\usepackage{gensymb}
\usepackage{bm}
\usepackage{graphicx}
\usepackage{pgfplots}
\pgfplotsset{compat=newest}
\newlength\figureheight
\newlength\figurewidth

\received{December 21, 2016}
\revised{February 13, 2017}
\accepted{February 17, 2017}
\shorttitle{Light Sail Stability}
\shortauthors{Manchester and Loeb}

\begin{document}

\title{Stability of a Light Sail Riding on a Laser Beam}

\author{Zachary Manchester\footnote{zmanchester@seas.harvard.edu}}
\affil{John A. Paulson School of Engineering and Applied Science, Harvard University \\ 60 Oxford St., Cambridge, MA 02138}

\author{Abraham Loeb\footnote{aloeb@cfa.harvard.edu}}
\affil{Astronomy Department, Harvard University\\ 60 Garden St., Cambridge, MA 02138}

\begin{abstract}
The stability of a light sail riding on a laser beam is analyzed both analytically and numerically. Conical sails on Gaussian beams, which have been studied in the past, are shown to be unstable without active control or additional mechanical modifications. A new architecture for a passively stable sail-and-beam configuration is proposed. The novel spherical shell design for the sail is capable of ``beam riding'' without the need for active feedback control. Full three-dimensional ray-tracing simulations are performed to verify our analytical results.
\end{abstract}

\section{Introduction}

The light sail concept---harnessing photon pressure to propel a spacecraft---has a long history dating back to some of the earliest pioneers of astronautics. Tsiolkovsky \& Zander first described ``tremendous mirrors of very thin sheets... using the pressure of sunlight to attain cosmic velocities'' in 1925 \citep{Zander1925}. Since then, most research has focused on solar sails---light sails that harness solar photons. Following the invention of lasers, in the 1960s Forward \citep{Forward1984}, Marx \citep{MarxNature}, and Redding \citep{Redding1967} independently proposed the use of high-power lasers to propel a sail to a significant fraction of the speed of light. This was followed by subsequent studies over the past five decades \citep{Moeckel1972, Weiss1979, Lubin}. Most recently, the \emph{Breakthrough Starshot Initiative}\footnote{http://breakthroughinitiatives.org/Initiative/3} was funded to propel a gram-scale spacecraft attached to a sail to a fraction of the speed of light using a high-power laser, with the goal of reaching the nearest stars within several decades.

There are many difficult engineering challenges associated with laser-propelled light sails that remain to be solved. A particularly important problem is ensuring that the sail remains centered on the laser beam despite disturbances, misalignment, and manufacturing imperfections. Ideally, a sail should possess \emph{beam-riding stability} without the need for active feedback control, as the addition of sensor and actuator hardware would add significant complexity and mass to the spacecraft.

While a substantial literature exists on the stability and control of solar sails \citep{Wie2004-1, Wie2004-2, Smith2005, Mimasu2011, Polites2008}, laser-propelled sails have received far less attention. The most closely related previous work has focused on conical microwave-propelled sails, which were studied both in numerical simulations \citep{Chahine2003, BenfordSim} and laboratory experiments \citep{BenfordExperiment1, BenfordExperiment2}. However, a rigorous theoretical analysis of the stability of such sails was not performed.

This paper analyzes the beam-riding stability of laser-propelled light sails and proposes a new passively stable laser and sail configuration. Section \ref{sec:BeamRiding} provides an introduction to the beam-riding problem, followed by a review of some basic results from linear stability theory in Section \ref{sec:Stability}. Next, Section \ref{sec:ConicalDynamics} derives a linearized dynamical model of a conical sail riding a Gaussian laser beam. Section \ref{sec:ConicalStability} then uses the model to show that such sail configurations are unstable without active feedback control or mechanical modifications. In Section \ref{sec:Spherical} we propose a novel passively stable spherical sail architecture. Section \ref{sec:Simulation} presents the results of numerical ray-tracing simulations that demonstrate the stability of the proposed design. Finally, Section \ref{sec:Discussion} summarizes our results and offers some commentary on future research directions.

\section{The Beam-Riding Problem} \label{sec:BeamRiding}

We assume that a laser beam with a full-width-at-half-maximum $W$ is incident on a sail of radius $R$, where $W$ and $R$ are of the same order of magnitude. The challenge is---through shaping the sail, choosing the beam profile, and possibly using active feedback control---to keep the sail centered on the beam as it is accelerated.

The total force applied by a beam incident on a perfectly reflective sail of area $S$ is given by,
\begin{equation} \label{eq:Force}
	\bm{F} = \int_{S} 2 \, \frac{P(\bm{x}) \, \hat{\bm{b}} \cdot \hat{\bm{n}}(\bm{x})}{c} \, \hat{\bm{n}}(\bm{x}) \, dS ,
\end{equation}
where the domain of integration is the surface of the sail, $\hat{\bm{n}}(\bm{x})$ is the unit vector normal to the sail surface at the point $\bm{x}$, $P(\bm{x})$ is the beam power flux at the point $\bm{x}$, $\hat{\bm{b}}$ is a unit vector parallel to the beam axis, and $c$ is the speed of light. Similarly, the total torque applied by the beam to the sail is given by,
\begin{equation} \label{eq:Torque}
	\bm{\tau} = \int_{S} 2 \, \frac{P(\bm{x}) \, \hat{\bm{b}} \cdot \hat{\bm{n}}(\bm{x})}{c} \, \big( \bm{r}(\bm{x}) \times \hat{\bm{n}}(\bm{x}) \big) \, dS,
\end{equation}
where $\bm{r}(\bm{x})$ is the vector from the sail's center of mass to point $\bm{x}$. If the sail is assumed to be rigid, its motion can be described by Newton's second law,
\begin{equation} \label{eq:Newton}
	m \ddot{\bm{x}} = \bm{F} ,
\end{equation}
and Euler's equation,
\begin{equation} \label{eq:Euler}
	I \dot{\bm{\omega}} + \bm{\omega} \times I \bm{\omega} = \bm{\tau} ,
\end{equation}
where $m$ is the mass, $\bm{\omega}$ is the angular velocity vector, and $I$ is the inertia tensor of the sail.

\section{Linear Stability Analysis} \label{sec:Stability}

We now briefly review some definitions and results from stability theory. For thorough treatments, the reader is referred to \citep{Kailath} and \citep{Khalil}.

A dynamical system can be generically written as a first-order vector differential equation,
\begin{equation} \label{eq:nonlinear}
	\dot{\bm{x}} = \bm{f}(\bm{x}),
\end{equation}
where $\bm{x} \in \mathbb{R}^n$ is the \emph{state vector} of the system. An \emph{equilibrium point} of the system is a point $\bm{x}^*$ such that,
\begin{equation}
	\bm{f}(\bm{x}^*) = 0 \,.
\end{equation}
Without loss of generality, we assume that $\bm{x}^*$ coincides with the origin.

A linear dynamical system is described by a square matrix $A \in \mathbb{R}^{n \times n}$ such that,
\begin{equation} \label{eq:linear}
	\dot{\bm{x}} = A \bm{x} \,.
\end{equation}
Nonlinear systems can be approximated in the neighborhood of the origin by taking $A$ to be,
\begin{equation}
	A_{ij} = \frac{\partial f_i}{\partial x_j} \,.
\end{equation}
Solutions of \eqref{eq:linear} are given by,
\begin{equation} \label{eq:soln}
	\bm{x}(t) = e^{At} \bm{x}_0 \,,
\end{equation}
where $\bm{x}_0$ is a vector of initial conditions and $e^{At}$ is a matrix exponential, which is formally defined in terms of its power series \citep{Kailath}.

The qualitative stability of a nonlinear system in the neighborhood of the origin is characterized by the eigenvalues of $A$. If all eigenvalues $\lambda_i$ have negative real parts, the matrix exponential in equation \eqref{eq:soln} will decay to zero as $t \to \infty$ and the state $\bm{x}(t)$ will tend toward the origin. In such cases, the system is said to be \emph{asymptotically stable}. On the other hand, if any eigenvalues have positive real parts, the exponential will grow unbounded as $t \to \infty$, and the system is said to be \emph{unstable}. Finally, if the real parts of any $\lambda_i$ are zero while the rest are negative, the system is said to be \emph{marginally stable}, and a definitive stability characterization cannot be made based on linearization \citep{Khalil}.

\section{Transverse Dynamics of Conical Sails on Gaussian Laser Beams} \label{sec:ConicalDynamics}

The notion of stability outlined in the previous section requires an equilibrium point. Clearly, the full dynamics of a beam-riding sail do not possess any equilibria: The sail accelerates as long as it remains on the beam. However, projecting the dynamics onto what we call the \emph{transverse subspace} results in a system with an equilibrium point at the origin.

\begin{figure}[h]
 \centering
% \ifdefined\SUBMIT
 	\includegraphics[width=3.5in]{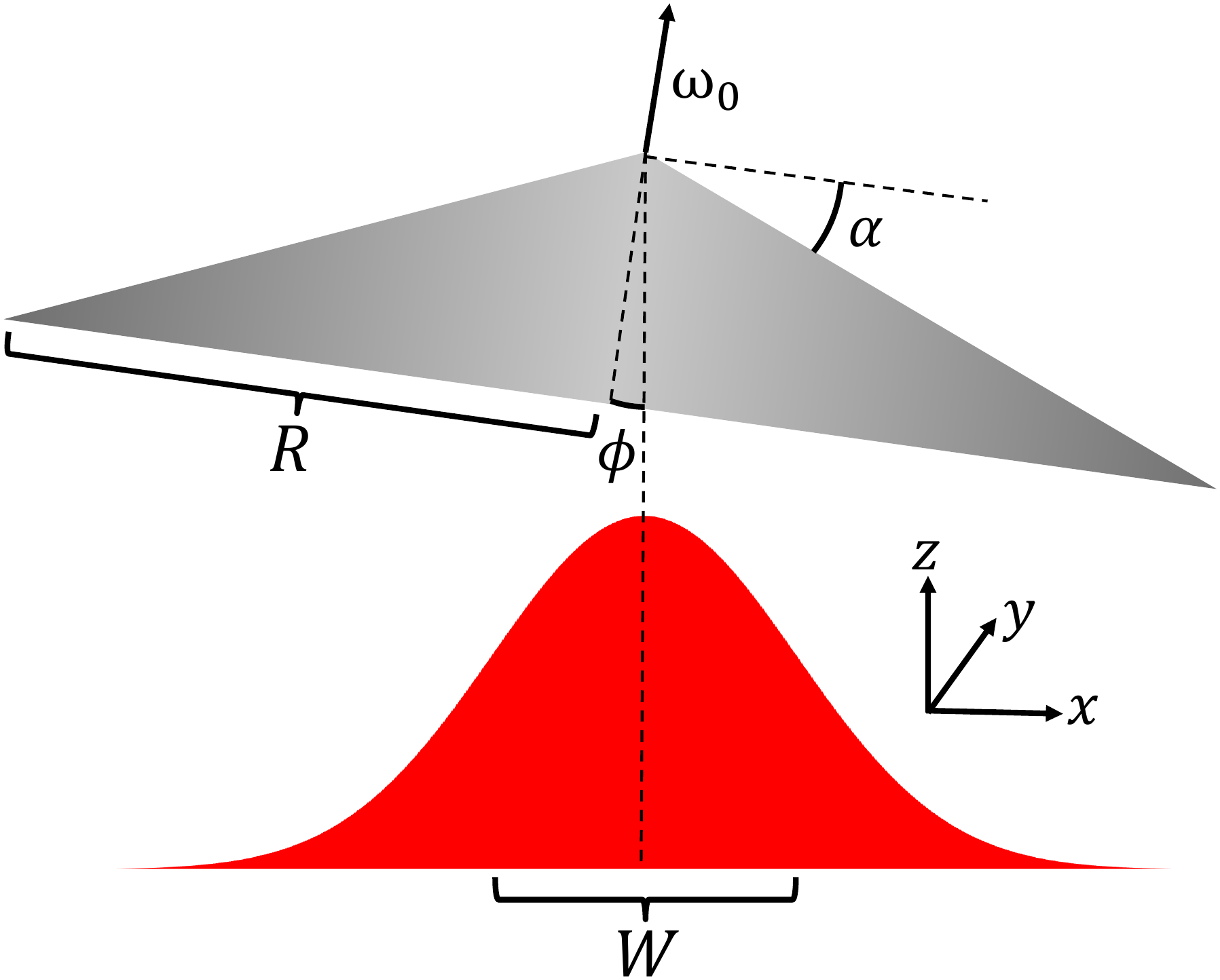}
% \else
% 	\includegraphics[width=3.2in]{diagram.pdf}
% \fi
 \caption{Schematic illustration of the geometry for a spinning conical sail riding on a Gaussian laser beam. \label{fig:diagram}}
\end{figure}

We define the transverse coordinates as those orthogonal to the beam axis. As depicted in Figure \ref{fig:diagram}, the coordinates $x$ and $y$ are used to describe translation of the sail in the plane orthogonal to the beam, while the angles $\theta$ and $\phi$ are used to describe rotation of the sail about the $x$ and $y$ axes, respectively. The $x$ and $y$ components of the angular velocity vector are denoted by $\omega_x$ and $\omega_y$. The laser beam is assumed to have a radially symmetric Gaussian power distribution whose width can be expressed in terms of the standard deviation $\sigma$ as,
\begin{equation}
	W = 2 \sigma \sqrt{2 \ln(2)} \,.
\end{equation}

In addition, we assume that the sail is symmetric about the $z$-axis, with mass $m$ and moments of inertia $I_z$ and $I_x = I_y$, that the sail is spinning about the $z$-axis with angular frequency $\omega_0$, and that the cone angle measured relative to the $x$-$y$ plane is $\alpha$. To simplify our analysis, we also assume that multiple reflections of the beam do not occur. For this to hold, $\alpha$ must be less than $30 \degree$.

We can now formally define beam-riding stability as the stability of the origin with respect to the sail's transverse dynamics. In the remainder of this section we derive a linear model that approximates these dynamics for a conical sail near the center of a Gaussian laser beam.

\subsection{Translation}

From basic geometry and ray optics, the translational motion of a conical sail near the center of the beam can be approximated by the equations,
\begin{equation} \label{eq:cone-force}
\begin{gathered}
	\ddot{x} = -k_1 x + k_2 \phi \\
	\ddot{y} = -k_1 y - k_2 \theta
\end{gathered} \,\,,
\end{equation}
where $k_1$ denotes the partial derivative of the sail's transverse acceleration with respect to displacements in the $x$-$y$ plane,
\begin{equation}
	k_1 = -\frac{1}{m} \frac{\partial F_x}{\partial x} = -\frac{1}{m} \frac{\partial F_y}{\partial y} \,,
\end{equation}
and $k_2$ is the partial derivative of the sail's transverse acceleration with respect to rotations about the $x$ and $y$ axes:
\begin{equation}
	k_2 = \frac{1}{m} \frac{\partial F_x}{\partial \phi} = -\frac{1}{m} \frac{\partial F_y}{\partial \theta} \,.
\end{equation}
The first term in equation \eqref{eq:cone-force} describes the restoring force due to the sail's conical shape, while the second term describes the forces encountered as the sail rotates by a small angle, causing a component of the beam to be deflected in the $x$-$y$ plane.

An expression for $k_1$ in terms of the system's parameters can be derived by integrating equation \eqref{eq:Force} over a sail with a small displacement $\delta_x$. In polar coordinates we have,
\begin{equation}
%\ifdefined\SUBMIT
	\Delta F_x = \int_0^R \int_0^{2 \pi} \frac{P_0}{\pi c \sigma^2} e^{-(r^2 + 2 \delta_x r \cos(\psi) + \delta_x^2)/2 \sigma^2} \cos(\alpha) \sin(\alpha) \cos(\psi) \, r \, dr \, d\psi \,,
%\else
%\begin{split}
%\Delta F_x = \int_0^R \int_0^{2 \pi} \frac{P_0}{\pi c \sigma^2} e^{-(r^2 + 2 \delta_x r \cos(\psi) + \delta_x^2)/2 \sigma^2} \dots \\
%\cos(\alpha) \sin(\alpha) \cos(\psi) \, r \, dr \, d\psi \,,
%\end{split}
%\fi
\end{equation}
where $\psi$ is the polar angle the standard deviation $\sigma$ is used in place of $W$ for clarity. Retaining only terms up to first order in $\delta_x$ gives:
\begin{equation}
%\ifdefined\SUBMIT
\begin{gathered}
	\Delta F_x = \int_0^R \int_0^{2 \pi} \frac{P_0}{\pi c \sigma^2} e^{-r^2/2 \sigma^2} \left( 1 + \frac{\delta_x r}{\sigma^2} \cos(\psi) \right) \cos(\alpha) \sin(\alpha) \cos(\psi) \, r \, dr \, d\psi \\
	= \frac{P_0 \delta_x}{2 c \sigma^4} \sin(2 \alpha) \int_0^R r^2 e^{-r^2/2 \sigma^2} \, dr \,.
\end{gathered}
%\else
%\begin{split}
%	\Delta F_x = \int_0^R \int_0^{2 \pi} \frac{P_0}{\pi c \sigma^2} e^{-r^2/2 \sigma^2} \left( 1 + \frac{\delta_x r}{\sigma^2} \cos(\psi) \right) \dots \\
%	\cos(\alpha) \sin(\alpha) \cos(\psi) \, r \, dr \, d\psi \\
%	= \frac{P_0 \delta_x}{2 c \sigma^4} \sin(2 \alpha) \int_0^R r^2 e^{-r^2/2 \sigma^2} \, dr \,.
%\end{split}
%\fi
\end{equation}
If $W \approx R$, most of the beam flux falls on the sail. Therefore we take the limit $R \to \infty$ to obtain the closed-form approximation,
\begin{equation}
	\Delta F_x \approx \frac{P_0 \sqrt{\pi} \sin(2 \alpha)}{c \sigma 2 \sqrt{2}} \delta_x \,.
\end{equation}
Finally, $\sigma$ is written in terms of $W$ to arrive at,
\begin{equation}
	k_1 = \frac{P_0 \sqrt{\pi \ln(2)}}{m c W} \sin(2 \alpha) \,.
\end{equation}

Performing similar steps, equation \eqref{eq:Torque} can be integrated over the surface of a sail rotated by a small angle to derive the following expression for $k_2$,
\begin{equation}
	k_2 = \frac{P_0}{m c} \left( 2\cos^2(\alpha) - \sin^2(\alpha) - \frac{d \sqrt{\pi \ln(2)}}{W} \sin(2 \alpha) \right) ,
\end{equation}
where $d$ is the distance from the tip of the cone to the sail's center of mass.

\subsection{Rotation}

The angular motion of the sail near the upright orientation $\theta = \phi = 0$ can be described by the kinematic equations,
\begin{equation} \label{eq:cone-angular}
\begin{gathered}
	\dot{\theta} = \omega_x + \omega_0 \phi \\
	\dot{\phi} = \omega_y + \omega_0 \theta \,,
\end{gathered}
\end{equation}
along with the following dynamics:
\begin{equation} \label{eq:cone-torque}
\begin{gathered}
	\dot{\omega}_x = -k_3 y + k_4 \theta -k_5 \omega_y \\
	\dot{\omega}_y = k_3 x + k_4 \phi + k_5 \omega_x \,.
\end{gathered}
\end{equation}

The constant $k_3$ describes the torque imparted by the beam on the sail due to translation in the $x$-$y$ plane. Proceeding in the same fashion as before, equation \eqref{eq:Torque} is integrated over the surface of a sail with a small displacement to arrive at,
\begin{equation} \label{eq:k2}
	k_3 = \frac{P_0}{c I_x} \left( 2 - \frac{d \sqrt{\pi \ln(2)}}{W} \sin(2 \alpha) \right) .
\end{equation}
The constant $k_4$ describes torques encountered as the sail rotates. Once again, we integrate equation \eqref{eq:Torque}---this time over the surface of a sail that has been rotated by a small angle---to derive:
\begin{equation}
%\ifdefined\SUBMIT
	k_4 = \frac{P_0}{c I_x} \left( d \left( 2 \cos(\alpha) - 3 \sin^2(\alpha) \right) - d^2 \frac{\sqrt{\pi \ln(2)}}{W} \sin(2 \alpha) - \frac{W \sqrt{\pi}}{4 \sqrt{\ln(2)}} \left( \sin(\alpha) - \sin^2(\alpha) \tan(\alpha) \right) \right) .
%\else
%\begin{split}
%	k_4 = \frac{P_0}{c I_x} \Bigg( d \left( 2 \cos(\alpha) - 3 \sin^2(\alpha) \right) - d^2 \frac{\sqrt{\pi \ln(2)}}{W} \sin(2 \alpha) \\
%	- \frac{W \sqrt{\pi}}{4 \sqrt{\ln(2)}} \left( \sin(\alpha) - \sin^2(\alpha) \tan(\alpha) \right) \Bigg) .
%\end{split}
%\fi
\end{equation}
Lastly, $k_5$ captures gyroscopic effects due to the sail's spin about the $z$-axis. It can be derived from Euler's equation \eqref{eq:Euler} by making the assumption that $\omega_0$ is much greater than both $\omega_x$ and $\omega_y$ \citep{Hughes}:
\begin{equation}
	k_5 = \omega_0 \left( \frac{I_x - I_z}{I_x} \right) .
\end{equation}

%Assembling equations \eqref{eq:cone-force}, \eqref{eq:cone-angular}, and \eqref{eq:cone-torque} into the generic form of equation \eqref{eq:linear} gives the following linear system:
%\begin{equation} \label{eq:cone-linear}
%	\begin{bmatrix}
%		\dot{x} \\ \dot{y} \\ \dot{\theta} \\ \dot{\phi} \\ \ddot{x} \\ \ddot{y} \\ \dot{\omega}_x \\ \dot{\omega}_y
%	\end{bmatrix} = 
%	\begin{bmatrix}
%		0 & 0 & 0 & 0 & 1 & 0 & 0 & 0 \\
%		0 & 0 & 0 & 0 & 0 & 1 & 0 & 0 \\
%		0 & 0 & 0 & \omega_0 & 0 & 0 & 1 & 0 \\
%		0 & 0 & -\omega_0 & 0 & 0 & 0 & 0 & 1 \\
%		-k_1 & 0 & 0 & k_2 & 0 & 0 & 0 & 0 \\
%		0 & -k_1 & -k_2 & 0 & 0 & 0 & 0 & 0 \\
%		0 & -k_3 & k_4 & 0 & 0 & 0 & 0 & k_5 \\
%		k_3 & 0 & 0 & k_4 & 0 & 0 & -k_5 & 0
%	\end{bmatrix}
%	\begin{bmatrix}
%		x \\ y \\ \theta \\ \phi \\ \dot{x} \\ \dot{y} \\ \omega_x \\ \omega_y
%	\end{bmatrix}.
%\end{equation}

\section{Stability of Conical Sails} \label{sec:ConicalStability}

Equations \eqref{eq:cone-force}, \eqref{eq:cone-angular}, and \eqref{eq:cone-torque} can be assembled into the matrix form,
\begin{equation}
	\dot{\bm{x}} = A_{\mathrm{cone}}\, \bm{x} \,.
\end{equation}
Directly calculating the eigendecomposition of $A_{\mathrm{cone}}$ analytically is quite unwieldy. Instead, we take advantage of its structure. First, we note that $A_{\mathrm{cone}}$ is traceless. Since the trace of a matrix is equal to the sum of its eigenvalues, the best that can be hoped for is to arrange all of the eigenvalues of $A_{\mathrm{cone}}$ to lie on the imaginary axis. It is important to keep in mind that such a marginal stability result does not allow us to make conclusions about the stability of the full nonlinear system.

\subsection{The Non-Spinning Case}

In the non-spinning case in which $\omega_0 = k_5 = 0$, the transverse sail dynamics can be written as a four-dimensional undamped oscillator:
\begin{equation} \label{eq:cone-oscillator}
	\ddot{\bm{x}} + K \bm{x} = 0 \,.
\end{equation}
In analogy with the scalar case, the ``spring constant'' matrix $K$ must have positive real eigenvalues for the system to be marginally stable. With $\bm{x} = \left[ x \,\, \theta \,\, y \,\, \phi \right]^T$, $K$ takes the following block-diagonal form:
\begin{equation}
	K = \begin{bmatrix}
		k_1 & -k_2 & 0 & 0 \\
		-k_3 & -k_4 & 0 & 0 \\
		0 & 0 & k_1 & k_2 \\
		0 & 0 & k_3 & -k_4
	\end{bmatrix} .
\end{equation}
Its eigenvalues can then be found in closed form by analyzing each $2 \times 2$ block separately:
\begin{equation} \label{eq:K_eigs12}
%\ifdefined\SUBMIT
\lambda_1 = \lambda_2 = \frac{1}{2} \Big( k_1 - k_4 + \sqrt{(k_1 - k_4)^2 + 4 k_1 k_4 + 4 k_2 k_3} \Big)
%\else
%\begin{aligned}
%	\lambda_1 = & \lambda_2 = \frac{1}{2} \Big( k_1 - k_4 \Big. \\
%	& \Big. + \sqrt{(k_1 - k_4)^2 + 4 k_1 k_4 + 4 k_2 k_3} \Big)
%	\end{aligned}
%\fi
\end{equation}
\begin{equation} \label{eq:K_eigs34}
%\ifdefined\SUBMIT
\lambda_3 = \lambda_4 = \frac{1}{2} \Big( k_1 - k_4 - \sqrt{(k_1 - k_4)^2 + 4 k_1 k_4 + 4 k_2 k_3} \Big) .
%\else
%\begin{aligned}
%	\lambda_3 = & \lambda_4 = \frac{1}{2} \Big( k_1 - k_4 \Big. \\
%	& \Big. - \sqrt{(k_1 - k_4)^2 + 4 k_1 k_4 + 4 k_2 k_3} \Big) .
%\end{aligned}
%\fi
\end{equation}

\begin{figure}[h]
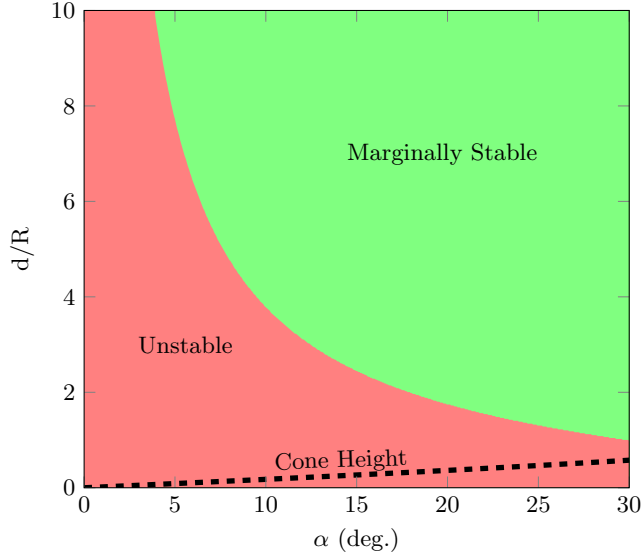

 \setlength{\figureheight}{2.5in}
 \setlength{\figurewidth}{3in}
 \centering
 % flatex input: [Stability.tikz]
% flatex input end: [Stability.tikz]

%\fi
 \caption{ \label{fig:stability}
  Stability regions in the space of sail parameters $\alpha$ (cone angle) and $d/R$ (center of mass location normalized by sail radius). The dashed line marks the normalized cone height.}
\end{figure}

From equation \eqref{eq:K_eigs34}, a necessary condition for marginal stability is,
\begin{equation} \label{eq:pendulum-inequality}
	k_1 k_4 + k_2 k_3 < 0 \,,
\end{equation}
which is plotted in Figure \ref{fig:stability} as a function of the cone angle, $\alpha$, and the distance between the cone tip and the center of mass $d$, which we have normalized by the cone radius $R$. A line showing the normalized cone height, which upper bounds $d/R$, is also plotted.

Figure \ref{fig:stability} indicates that a simple conical sail cannot achieve stability, since the center of mass must lie beneath the base of the cone by a significant distance. While it may appear that further increasing the cone angle could alleviate this problem, any benefit will be very limited since the restoring force on the sail begins to decrease beyond $\alpha = 30 \degree$ due to multiple reflections of the beam. In fact, it is easy to show that there is no restoring force when $\alpha = 45 \degree$, indicating instability. One can imagine mechanical solutions that could lower the center of mass, such as the rigid pendulum suggested by \citep{Chahine2003}, however, they introduce serious practical difficulties. First, any such structure will necessarily be exposed to the laser beam. Second, significant additional mass would be added to the spacecraft, reducing its acceleration. Lastly, the flexible modes of the structure and their effect on stability would require careful analysis.

\subsection{The Spinning Case}

We now turn to the spinning case in which $\omega_0$ and $k_5$ are non-zero. First, we note that it is possible to achieve marginal stability with a sufficiently large choice of $\omega_0$. However, the situation is somewhat more subtle than might be expected from an analysis of the linearized transverse dynamics.

While equations \eqref{eq:cone-angular} and \eqref{eq:cone-torque} capture gyroscopic precession and nutation effects, they implicitly assume that the sail's angular momentum vector $\bm{\ell} = I \bm{\omega}$ is perfectly aligned with the beam axis $\hat{\bm{b}}$. If the sail has an initial $\bm{\ell}$ that is not in perfect alignment with the beam axis, the $\mathrm{SO}(2)$ symmetry of the system is broken and the equilibrium point in the transverse dynamics \emph{disappears}. A stability analysis in the sense of Section \ref{sec:Stability} is therefore misleading.

\begin{figure}[h]
 \centering
 \includegraphics[width=3.5in]{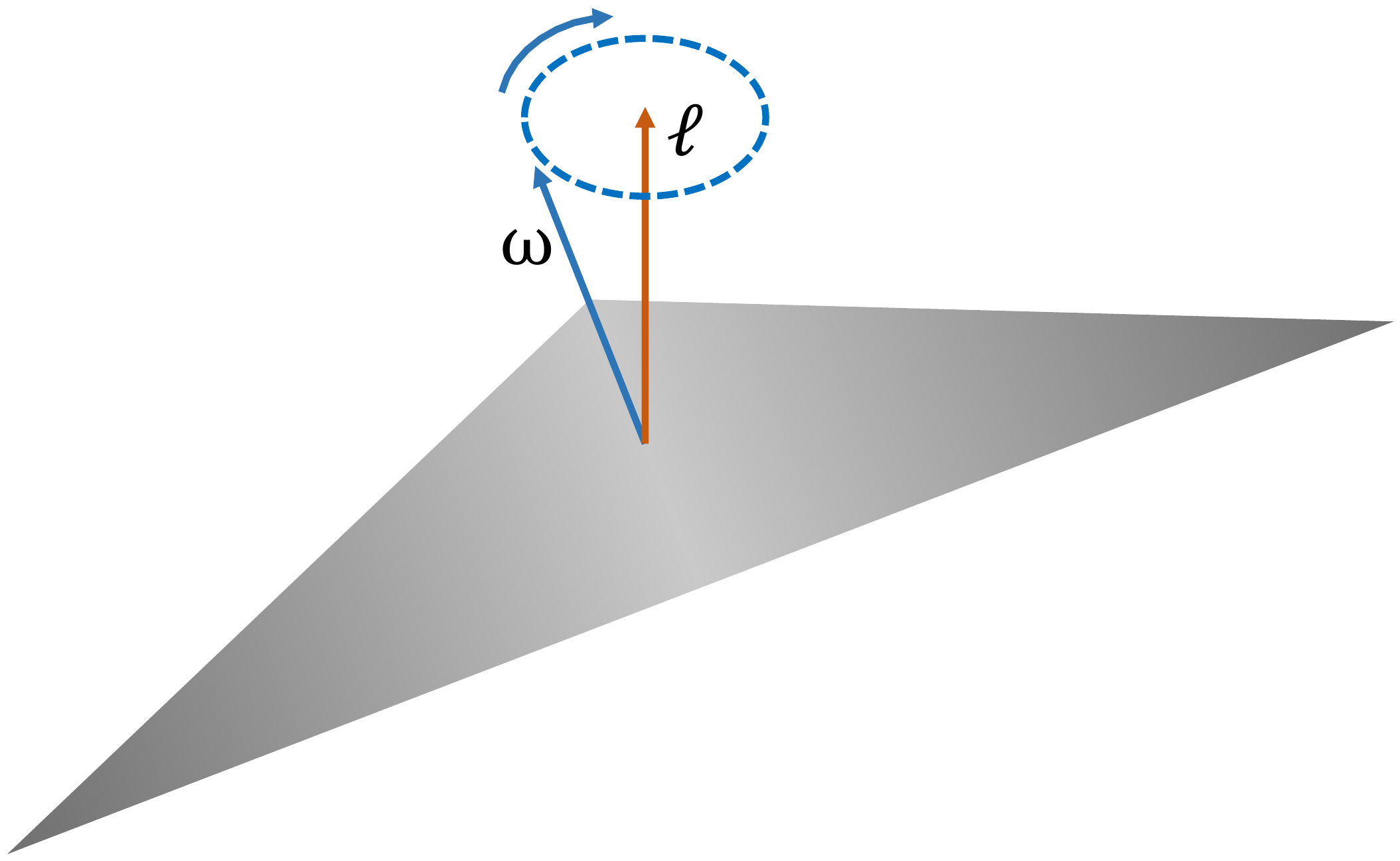}
 \caption{Illustration of the effect of gyroscopic precession on the angular velocity of a sail with angular momentum $\bm{\ell}$. \label{fig:precession}}
\end{figure}

A qualitative physical understanding of the situation can be gained by recalling the behavior of a rigid body undergoing precession. Viewed in an inertial reference frame, the body's angular velocity vector $\bm{\omega}$ traces out a cone centered on its angular momentum vector (Figure \ref{fig:precession}). In the case of a conical sail with $\bm{\ell}$ parallel to the beam axis, one can see that it should be possible, with a sufficiently high spin frequency $\omega_0$, for perturbing forces to ``average out'' over a precession period. However, if $\bm{\ell}$ is not \emph{exactly} parallel to the beam axis, the average force on the sail over a precession period will have a component in the $x$-$y$ plane, pushing the sail off the beam.

In mathematical terms, the equilibrium point in the sail's transverse dynamics is not \emph{structurally stable}. It exists only for very particular choices of $\bm{\ell}$, the set of which has zero Lebesgue measure. In practical terms, it is not possible to achieve perfect alignment of the sail's angular momentum vector with the beam axis. Therefore, we cannot expect a spinning conical sail to achieve stable beam riding.

\section{Spherical Shell Design for the Sail} \label{sec:Spherical}

We now propose an alternative beam-riding architecture with more favorable stability properties. Upon inspection of the matrix $A_{\mathrm{cone}}$, it is clear that the instabilities found in the previous section are rooted in coupling between the translation and rotation degrees of freedom of the sail. Motivated by this observation, we analyze a spherical shell configuration for the sail, whose symmetry eliminates such coupling.

\begin{figure}[h]
 \centering
 \includegraphics[height=2.5in]{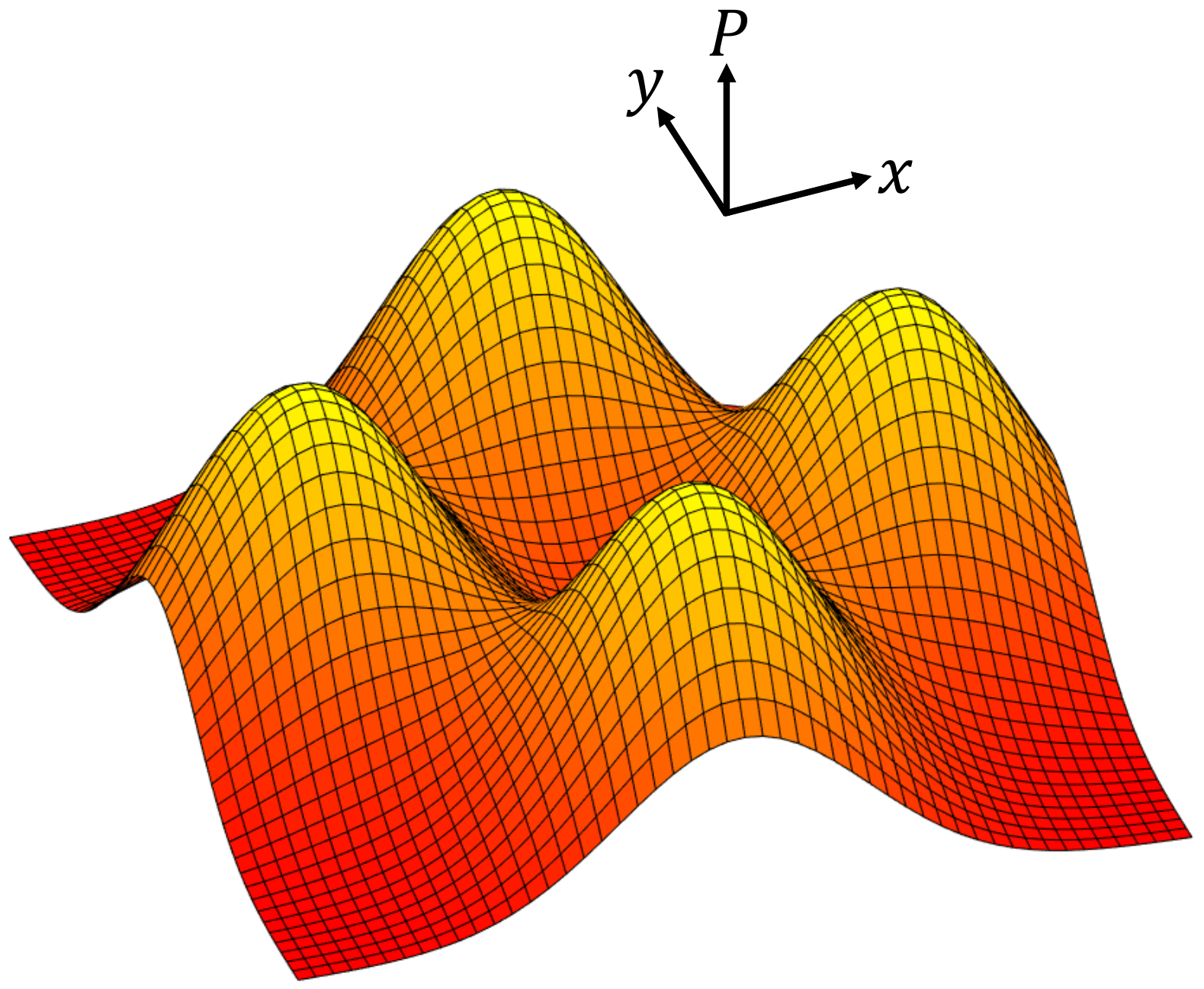}
 \includegraphics[height=2.3in]{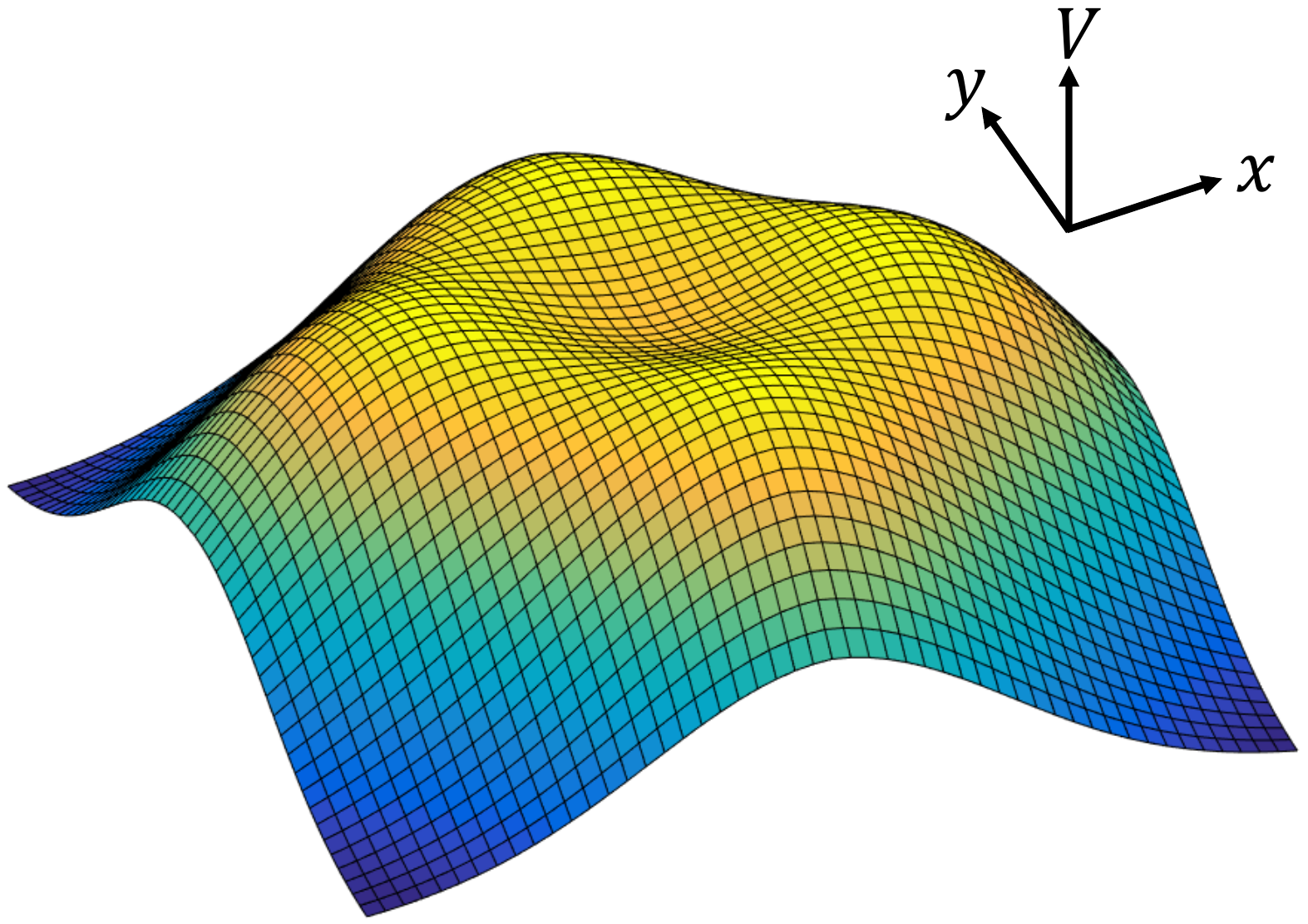}
 \caption{ \label{fig:multimodal}
  Multimodal beam profile composed of four Gaussian laser beams (left) and corresponding potential function $V(x,y)$ for the transverse dynamics of a spherical sail (right).}
\end{figure}

A light ray offset from the centerline of a reflective spherical sail clearly does not produce a restoring force, instead pushing the sphere farther away from the beam axis. Any unimodal beam profile, like the Gaussian studied in the previous section, will have a similar effect. If the beam is instead allowed to be multimodal, stable beam riding becomes possible. The left side of Figure \ref{fig:multimodal} depicts a beam profile composed of a sum of four Gaussians. A sum of Gaussians was chosen due to the ease of producing Gaussian beams and their favorable divergence properties. An appropriately sized spherical sail perturbed from the center of such a composite beam will experience a restoring force pushing it back toward the center due to the increased power in the sides of the beam.

The linearized transverse dynamics for this beam riding configuration are,
\begin{equation} \label{eq:sphere-force}
\begin{aligned}
	& \ddot{x} = -k_1 x & \ddot{\theta} = 0 \\
	& \ddot{y} = -k_1 y & \ddot{\phi} = 0 \,,
\end{aligned}
\end{equation}
where, once again,
\begin{equation}
	k_1 = -\frac{1}{m} \frac{\partial F_x}{\partial x} = -\frac{1}{m} \frac{\partial F_y}{\partial y} .
\end{equation}
As expected, there is no coupling between translation and rotation. There are also no torques applied to the sphere by the beam, since all forces are directed through the center of mass. By inspection, the associated eigenvalues are $0$ and $\pm i \sqrt{k_1}$. Since all lie on the imaginary axis, the system is marginally stable.

To obtain a more conclusive stability result, the beam is discretized on a grid and the forces on the sail are evaluated at each grid point. Since the transverse forces $\bm{F}_{\perp}$ depend only on the sail's position in the $x$-$y$ plane and are not functions of velocity or time, they are conservative \citep{Goldstein}. As a result, they can be associated with a scalar potential function $V(x,y)$ such that $\bm{F}_{\perp} = -\nabla V(x,y)$. We compute this potential function numerically, as depicted in the right half of Figure \ref{fig:multimodal}.

Based on Figure \ref{fig:multimodal}, it is clear that there is a basin of attraction surrounding the center of the beam. As long as the sail's initial conditions lie within this basin and its total energy is below the energy associated with the rim of the basin (which can be calculated numerically for any parameters of interest), the sail will remain trapped in the basin. Physically, the sail will oscillate around the center of the beam, but the amplitude of the oscillations will remain bounded. While this does not meet the definition of asymptotic stability presented in Section \ref{sec:Stability}, it does meet the looser requirements of \emph{Lyapunov stability} \citep{Khalil}.

Figure \ref{fig:multimodal} also implies an inherent trade-off between stability and the acceleration experienced by the sail along the $z$-axis. By placing the beam's constituent Gaussians closer together, more flux will fall onto the sail and it will experience greater acceleration. However, the size of the potential well will also be reduced, making it easier for the sail to be pushed off the beam. This trade-off manifests itself in some form in all sail designs. Ultimately, an evaluation must be made based on the size of the perturbations encountered in practice.

\section{Numerical Simulations} \label{sec:Simulation}

We demonstrated the stability of the spherical sail riding on the composite beam profile shown in Figure \ref{fig:multimodal} in two numerical simulations. The integrals in equations \eqref{eq:Force} and \eqref{eq:Torque} were approximated by discretizing the beam shown in Figure \ref{fig:multimodal} into a grid of $50 \times 50$ rays. The path of each ray was then traced as it intersected the sail and reflected off of its surface. The net change in momentum of each ray was calculated, and the resulting forces and torques were applied to the sail. The differential equations \eqref{eq:Newton} and \eqref{eq:Euler} were then integrated forward in time using the standard fourth-order Runge-Kutta method. The parameters used in our simulations follow the Starshot design, with a beam power $P = 100$ GW, a sail mass $m = 10$ g, a sphere radius $R = 1$ m, a width of each constituent Gaussian in the beam of $W = 1$ m, and a distance of 1 m between the center of each constituent Gaussian and the overall beam center.

\begin{figure}[h]
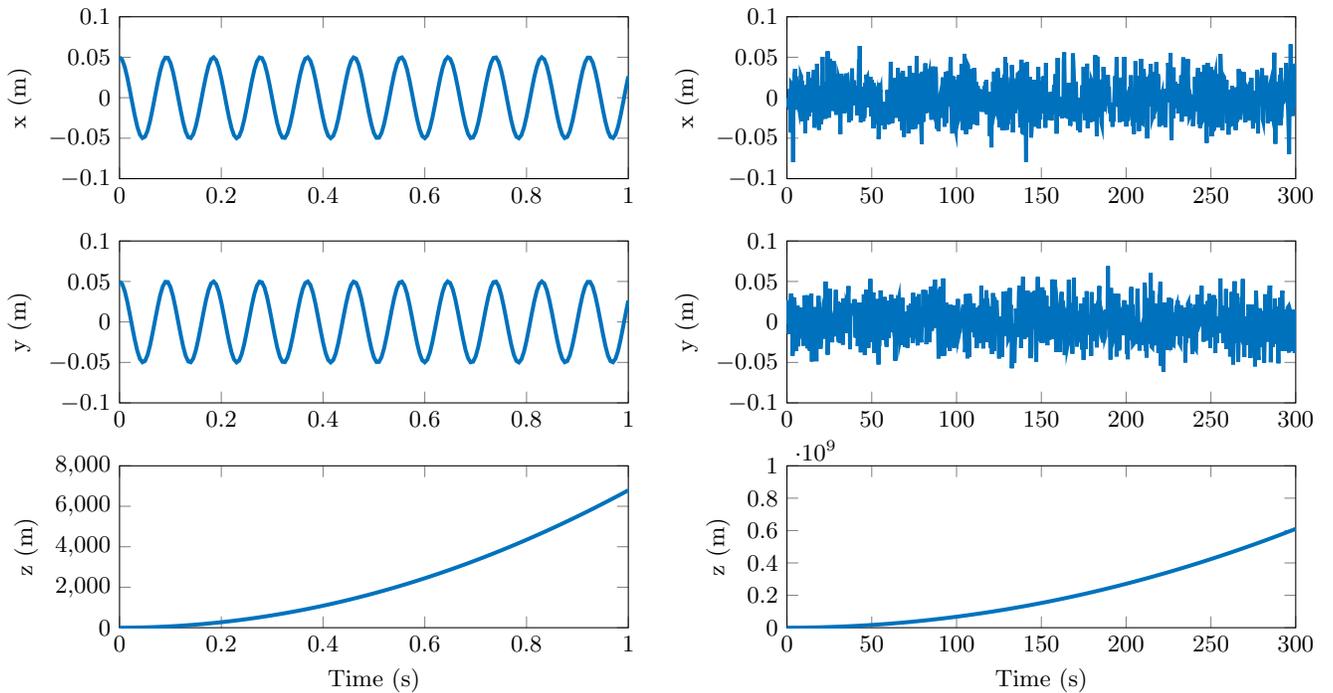

 \setlength{\figureheight}{3.2in}
 \setlength{\figurewidth}{2.8in}
 \centering
  % flatex input: [Sim1.tikz]
\definecolor{mycolor1}{rgb}{0.00000,0.44700,0.74100}%
%
% flatex input end: [Sim2.tikz]

%\fi
 \caption{ \label{fig:sim}
  Sail position during beam-riding simulations without (left) and with (right) noise added to the beam.}
\end{figure}

The left side of Figure \ref{fig:sim} shows the components of the sail's position vector during a short simulation with an initial offset of 5 cm in both the $x$ and $y$ components of the position vector and zero initial velocity. The sail's position in the $x$-$y$ plane oscillates with a frequency of roughly 11 Hz but, as predicted, remains bounded. 

The right side of Figure \ref{fig:sim} shows the components of the sail's position vector during a longer simulation in which white noise was added to the rays making up the laser beam to simulate perturbations due to atmospheric turbulence. The average power of the noise applied to each ray was chosen to be 1\% of that ray's nominal power. While perturbations of the beam clearly excite transverse oscillatory motion, the sail remains in the stable basin of attraction over a time scale of several minutes, which is sufficient for it to achieve a sizable fraction of the speed of light along the $z$-axis.

In general, noise will add energy to the transverse modes of the system. This energy will execute a random walk and, after sufficient time, will exceed the energy associated with the rim of the potential well, causing the sail to leave the beam. This ``exit time,'' which depends on the beam power and shape, as well as the power spectral density of the noise, will be an important consideration in the design of a realistic laser-sail system.

\section{Discussion} \label{sec:Discussion}

We have presented a passively stable laser-sail architecture that is capable of beam riding without active feedback control. The proposed design makes use of a spherical sail and a multimodal beam profile. While we have focused on a particular beam profile composed of a sum of four Gaussians, many others are possible, including sums of three or more Gaussians and radially symmetric ring-like profiles.

Spherical sails possess a number of practical advantages over the conical sail designs previously studied in the literature. First, a hollow spherical sail could be made of a thin, flexible material and inflated with a gas to maintain its shape. This would allow many sails to be stored compactly and launched inside a conventional rocket before being inflated in Earth orbit and accelerated with a ground-based laser. It would also allow the sail to be deflated after the acceleration phase, and perhaps reconfigured for some other purpose. A spherical sail would also be much less massive than a rigid conical sail with a pendulum, allowing it to achieve a higher acceleration with a given beam power. Lastly, the interior of a spherical sail would naturally provide a shielded environment for electronics or other payload items.

There are several effects that were not accounted for in this study, but which are likely to be important in the practical implementation of a laser-sail system. Perhaps most importantly, we have assumed a perfectly rigid sail. In practice, the sail will have flexible structural modes which may impact its beam-riding dynamics. Deformation of the sphere's surface could cause non-zero torques on the sail, but could also provide damping, which would reduce the amplitude of oscillations about the center of the beam. It may also be possible to use actuators to actively adjust the sail's stiffness and damping properties.

\section*{Acknowledgements}

The authors acknowledge support from the Breakthrough Prize Foundation, and are grateful to the Starshot group at Harvard for comments on the content of this paper.

%\ifdefined\SUBMIT
	%\bibliography{beam-rider.bib}

%\else
%	\bibliography{beam-rider.bib}
%\fi

\end{document}